\documentclass[a4paper,fleqn,usenatbib,letters]{mnras}

\usepackage[T1]{fontenc}
\usepackage{ae,aecompl}

\usepackage{graphicx}	% Including figure files
\usepackage{amsmath}	% Advanced maths commands
\usepackage{amssymb}	% Extra maths symbols

\title[Energy budget of stellar magnetic fields]{The energy budget of stellar magnetic fields: comparing non-potential simulations and observations}

\author[L. T. Lehmann et al.]{
L. T. Lehmann,$^{1}$\thanks{E-mail: ltl@st-andrews.ac.uk}
M. M. Jardine,$^{1}$
A. A. Vidotto,$^{2}$
D. H. Mackay,$^{3}$ 
V. See,$^{1}$ \newauthor
\,J.-F. Donati,$^{4,5}$
C. P. Folsom,$^{4,5}$
S. V. Jeffers,$^{6}$
S. C. Marsden,$^{7}$
J. Morin,$^{8}$
P. Petit$^{4,5}$
\\
$^{1}$School of Physics and Astronomy, University of St Andrews, St Andrews KY16 9SS, UK \\
$^{2}$School of Physics, Trinity College Dublin, The University of Dublin, Dublin-2, Ireland\\
$^{3}$School of Mathematics and Statistics, University of St Andrews, St Andrews KY16 9SS, UK\\
$^{4}$Universit\'e de Toulouse, UPS-OMP, IRAP, Toulouse, France\\
$^{5}$CNRS, Institut de Recherche en Astrophysique et Planetologie, 14 Avenue Edouard Belin, F-31400 Toulouse, France\\
$^{6}$Institut f\"ur Astrophysik, Georg-August-Universit\"at G\"ottingen, Friedrich-Hund-Platz 1, 37077, G\"ottingen, Germany\\
$^{7}$Computational Engineering and Science Research Centre, University of Southern Queensland, Toowoomba 4350, Australia\\
$^{8}$LUPM, Universit\'e de Montpellier, CNRS, Place Eug\`ene Bataillon, 34095, France\\
}

\date{Accepted XXX. Received YYY; in original form ZZZ}

\pubyear{2016}

\begin{document}
\label{firstpage}
\pagerange{\pageref{firstpage}--\pageref{lastpage}}
\maketitle

% Abstract of the paper
\begin{abstract}
The magnetic geometry of the surface magnetic fields of more than 55 cool stars have now been mapped using spectropolarimetry. In order to better understand these observations, we compare the magnetic field topology at different surface scale sizes of observed and simulated cool stars. For ease of comparison between the high-resolution non-potential magnetofrictional simulations and the relatively low-resolution observations, we filter out the small-scale field in the simulations using a spherical harmonics decomposition. 
We show that the large-scale field topologies of the solar-based simulations produce values of poloidal/toroidal fields and fractions of energy in axisymmetric modes that are similar to the observations. These global non-potential evolution model simulations capture key magnetic features of the observed solar-like stars through the processes of surface flux transport and magnetic flux emergence. They do not, however, reproduce the magnetic field of M-dwarfs or stars with dominantly toroidal field. Furthermore, we analyse the magnetic field topologies of individual spherical harmonics for the simulations and discover that the dipole is predominately poloidal, while the quadrupole shows the highest fraction of toroidal fields. Magnetic field structures smaller than a quadrupole display a fixed ratio between the poloidal and toroidal magnetic energies.
\end{abstract}

% Select between one and six entries from the list of approved keywords.
% Don't make up new ones.

%%%%%%%%%%%%%%%%%%%%%%%%%%%%%%%%%%%%%%%%%%%%%%%%%%

%%%%%%%%%%%%%%%%% BODY OF PAPER %%%%%%%%%%%%%%%%%%

\section{Introduction}

% Fig 1: Surface Maps
\begin{figure*}
	\includegraphics[width=\textwidth]{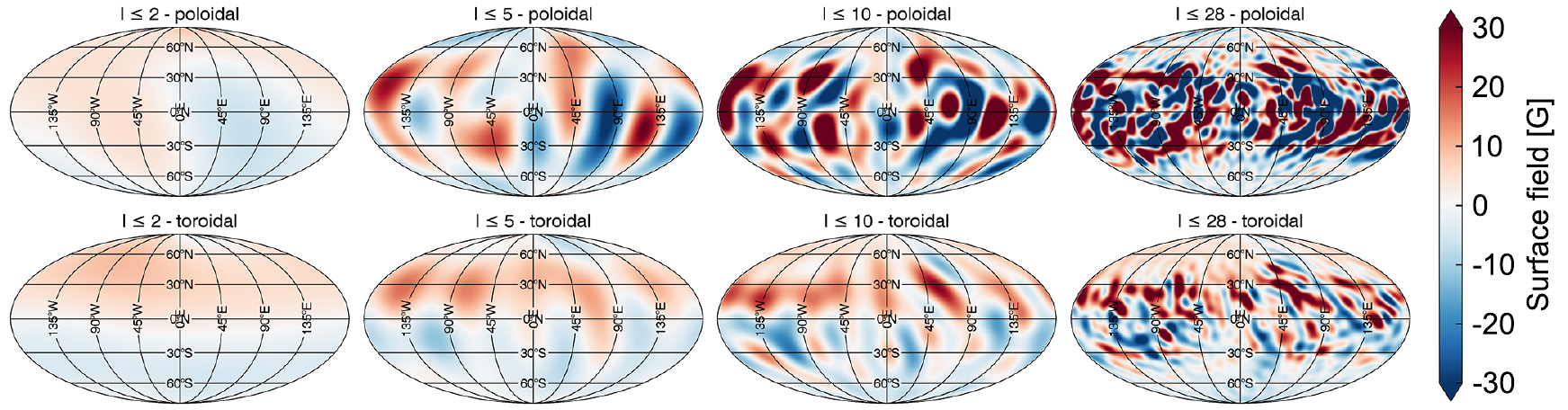}
    \caption{The surface magnetic field for a simulated star with three times the solar flux emergence rate and three times the solar differential rotation modulated with a flux transport model \citep{Gibb2016} restricted to spherical harmonic $\ell$-modes of $\ell \leq 2$ (left), $\ell \leq 5$ (middle left), $\ell \leq 10$ (middle right), and for $\ell \leq 28$ (right). The top row displays the poloidal and the bottom row the toroidal field component. The main polarity pattern of the toroidal field of the emerging bipoles, i.e., the polarity reversal across the equator, can be detected through all $\ell$-modes down to $\ell \leq 2$. The colourbar saturates at $\pm 30$\,G. }
    \label{fig:SurfaceMaps}
\end{figure*}

The magnetic activity of solar-like stars underpins both their high-energy coronal emission and also their angular momentum evolution through the action of their hot, magnetically-channelled winds. Our understanding of this activity is based on what we have learned about the Sun.
Solar activity phenomena, e.g., sunspots, prominences, and coronal holes, are mainly driven by the Sun's magnetic field and show a cyclic behaviour. During the activity cycle, the large-scale field of the Sun develops from an axisymmetric dipole (see, e.g. \citealt{Ossendrijver2003}), to a more chaotic small-scale structured field, covering mid to low latitudes and then back to a reversed dipole. \cite{DeRosa2012} confirmed that the dipole component follows the activity cycle in anti-phase, while the quadrupole component is in phase with the activity cycle in a similar way to the spot coverage. While the magnetic flux density in spots may be several thousand Gauss, however, the global solar magnetic flux density is only a few Gauss \citep{Babcock1955, Mancuso2007}.

It was \cite{Larmor1919} who first suggested that the solar magnetic field is induced by plasma motions. This led to the first dynamo model \citep{Parker1955}: the weak poloidal field is wound around the star by differential rotation ($\Omega$-effect) creating a strong toroidal band, while cyclonic turbulence restores the poloidal field from this toroidal field with opposite polarity ($\alpha$-effect). After that, different dynamo models were developed, such as the flux transport model by \cite{Babcock1961} and \cite{Leighton1969}, which can reproduce many observed solar behaviours (see also review by \citealt{Mackay2012}). This model predicts the surface magnetic flux that results from the injection of bipolar sunspot pairs which then undergo shearing by differential rotation, poleward meridional flow, and diffusion (\citealt{Wang1989, Baumann2004, Mackay2004, Jiang2013}). 
Recently, \cite{Gibb2016} studied the influence of two stellar parameters, namely differential rotation and flux emergence rate, on the non-potential coronal field by applying a flux transport model based on solar observations \citep{Mackay2006, Yeates2012}. They found that the flux emergence rate strongly influences the magnetic field properties and that an increased differential rotation opens the coronal magnetic field and makes it more non-potential. 

To obtain a wider understanding of the solar dynamo, the analysis of other cool stars magnetic fields is essential. The Zeeman-Doppler-Imaging technique (ZDI, \citealt{Donati1997, Donati2006a}) enables us to study the large-scale magnetic field topology (intensity and orientation of field distributions across the stellar surface) by analysing sequences of circularly polarised spectra (Stokes~V). ZDI can determine the contribution of the different field components (poloidal/toroidal, axisymmetric/non-axisymmetric, dipole/quadrupole/higher multi-poles) but only for the large-scale field as the ZDI technique suffers from the cancellation of opposite magnetic polarities on smaller scales.
In contrast, Zeeman broadening measurements use unpolarised light (Stokes~I) and are sensitive to the total magnetic flux emerged at the surface of stars but have little to no ability to document the field topology at any scale \citep{Robinson1980, Saar1988, Reiners2006, Lehmann2015a}.

Several ZDI surveys of different kinds of cool stars have exposed a wide range of magnetic field topologies, (e.g. \citealt{Donati2006, Marsden2006, Petit2008, Morin2010, Fares2013}). 
\cite{Vidotto2014} analysed how the large scale magnetic field of cool stars evolves with age, rotation period, Rossby number and X-ray luminosity. Recently, \cite{Folsom2016} added a new set of young solar type stars.
\cite{See2015} analysed the magnetic field topologies of 55 stars with masses in the range of 0.1-1.5$\,\mathrm{M_{\odot}}$ and discovered that the toroidal field scales with the inverse Rossby number more steeply than the poloidal field does. The toroidal field fraction shows two different power law dependences on the poloidal field for stars above and below 0.5$\,\mathrm{M_{\odot}}$. Additionally, they found that strong toroidal fields are typically axisymmetric.

To set highly-resolved solar or simulated magnetic vector maps into the context of ZDI-observed cool stars, \cite{Vidotto2016a} presented a magnetic field decomposition method. The method is compatible to the description used in ZDI-studies and decomposes the magnetic field into spherical harmonics of different $\ell$-modes up to a maximum $\ell_{\mathrm{max}}$. The small $\ell$-modes describe the large-scale field ($\ell=1$ dipolar, $\ell=2$ quadrupolar mode, etc.) and the higher $\ell$-modes the small-scale field. This small-scale field can be removed from the simulations or solar observations by selecting low $\ell$-modes to compare their magnetic field topology with the magnetic field topology of observed stars. 

In this paper we compare the global time-dependent non-potential simulations of \cite{Gibb2016} based on the surface flux transport model, with the sample of observed stars\footnotemark analysed by \cite{See2015}. In particular, we compare their magnetic field topologies, focusing on the poloidal, toroidal, and their axisymmetric components at different spatial scales. 
\footnotetext{The observations including results from the Bcool and Toupies survey were published by Petit (in prep.); \cite{BoroSaikia2015, doNascimento2014, Donati2003, Donati2008, Fares2009, Fares2010, Fares2012, Fares2013, Folsom2016, Morin2008a, Morin2008, Morin2010, Jeffers2014,  Petit2008, Waite2011}.}

\section{Simulations and Methods}
\label{sec:SimulationsandMethods}

The simulations of \cite{Gibb2016} used a magnetic flux transport model for the photospheric evolution, coupled with a non-potential coronal evolution model described by a magnetofrictional technique. The magnetic flux transport model solves the magnetohydrodynamic induction equation to evolve the surface magnetic flux, whilst a flux emergence pattern, e.g., bipolar starspot pairs, are advectively injected, and then sheared by surface differential rotation, poleward meridional flow, and diffusion \citep{Charbonneau2014}.
The flux emergence pattern is based on \cite{Yeates2014} who determined flux emergence properties using solar synoptic magnetograms observed by the US National Solar Observatory, Kitt Peak, between 2010/01 and 2011/01. \cite{Gibb2016} then varied the flux emergence rate (ER) using a statistical model, and the differential rotation (DR) in the range of $1\leq \mathrm{ER/ER_{\odot}} \leq 5$ and $1 \leq \mathrm{DR/DR_{\odot}} \leq 5$. The variation of the parameters leads us to 17 simulations ranging from solar-like stars up to stars with five times solar flux emergence rate and differential rotation. 

The immense difference in resolution prevents a direct comparison between the highly-resolved simulated and the relatively poorly-resolved observed stellar magnetic field vector maps. \cite{Vidotto2016a} presented a method to filter out the small-scale field from highly-resolved solar magnetic field vector maps to ease the comparison with stellar magnetic field vector maps. The method decomposes the radial, azimuthal, and meridional magnetic field components into their spherical harmonics of different $\ell$-modes. This enables us to investigate the magnetic field topology on different surface scale sizes ($\theta \approx 180^{\circ}/\ell$), see Fig.~\ref{fig:SurfaceMaps}. The usage of only the low-order $\ell$-modes, e.g. $\ell \le 10$, recovers the large-scale magnetic field topology and secures a fair order of magnitude comparison with most observed stellar magnetic field topologies. 

The stellar magnetic field geometry is often described by its poloidal/toroidal and its axisymmetric component, e.g. \cite{Donati2006, Petit2008, Donati2009, See2015, Vidotto2016}. We can determine these components by decomposing the magnetic field vectors ($B_r, B_{\theta}, B_{\phi}$) into their spherical harmonics after \cite{Elsasser1946} and \citet[Appendix III]{Chandrasekhar1961}. As described in Eq.~(2)-(8) of \cite{Donati2006a}, the poloidal field is characterised by the coefficients $\alpha_{\ell m}$ and $\beta_{\ell m}$ and the toroidal field by $\gamma_{\ell m}$, see Appendix~A. We define axisymmetric modes by $m=0$, thus including only exactly aligned modes. 
As the ZDI technique is also based on a spherical harmonic decomposition, we apply the same procedure to decompose both the simulated and observed magnetic field vector maps into their poloidal/toroidal and axisymmetric components.

\section{Comparing magnetic field topologies of the simulated and observed stars}

% Fig 2: B_tor^2 vs B_pol^2
\begin{figure}
	\includegraphics[width=\columnwidth]{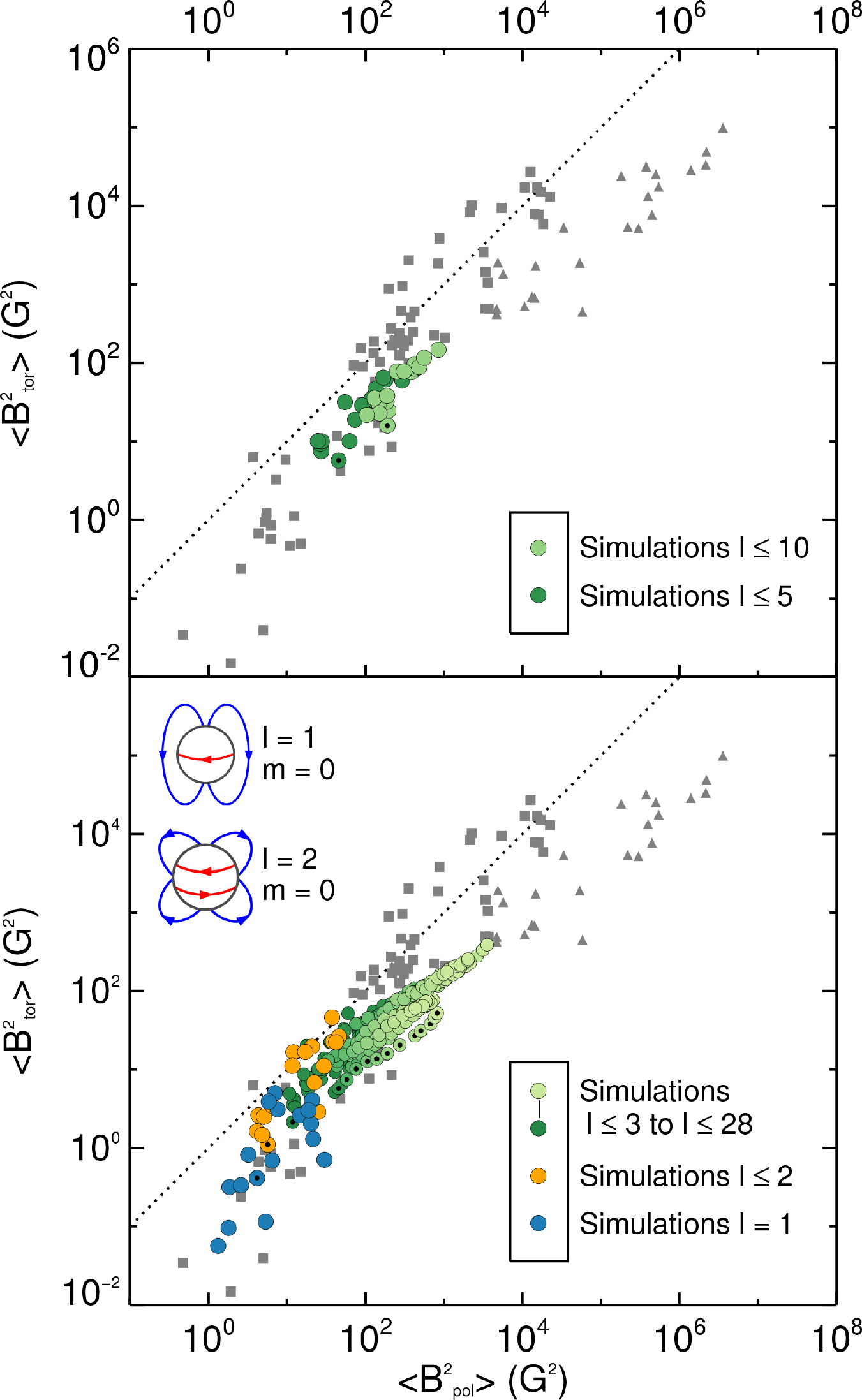}
    \caption{Magnetic field energy stored in the toroidal $\langle B^2_{\mathrm{tor}}\rangle$ and poloidal fields $\langle B^2_{\mathrm{pol}}\rangle$. Results of simulations are shown as coloured circles, while those from observations are shown as grey symbols, where stars with masses equal or above 0.5$\,\mathrm{M_{\odot}}$ are plotted as squares and stars with masses lower than 0.5$\,\mathrm{M_{\odot}}$ as triangles. The simulation representing the Sun is marked by the solar symbol $\odot$. The dashed line indicates equal toroidal and poloidal energies. \textit{Top:} The simulations are restricted to the large-scale field by spherical harmonics up to $\ell \leq 5$ (dark green circles) or $\ell \leq 10$ (light green circles) for a reasonable comparison with the observations. \textit{Bottom:} The simulations for all surface scales sizes: the dipolar component $\ell=1$ (blue circles), the quadrupolar component $\ell \leq 2$ (orange circles), and the higher $\ell$-modes $\ell \leq 3$ to $\ell \leq 28$ (greenish circles), where the colour gets lighter with increasing $\ell$-mode. The inserts show the poloidal (blue) and toroidal (red) field lines for the axisymmetric  dipole and quadrupole mode.}
    \label{fig:BtorBpol}
\end{figure}

\subsection{Magnetic energy stored in the toroidal/poloidal field}
\label{subsec:torpol}

% Fig. 2 top
We compare the magnetic energy budgets of the vector magnetic field from the observed stars studied by \cite{See2015} with the photospheric vector magnetic field from the simulated stars modelled by \cite{Gibb2016}. Firstly, we focus on the toroidal and poloidal magnetic field component (see Fig.~2~(top), \citealt{See2015}). In Figure~\ref{fig:BtorBpol} we plot the mean squared magnetic flux density of the toroidal field $\langle B^2_{\mathrm{tor}}\rangle$ against the poloidal field $\langle B^2_{\mathrm{pol}}\rangle$, where $\langle B^2_{\mathrm{pol/tor}}\rangle = \tfrac{1}{4\pi}\textstyle \int \sum_k \textstyle B^2_{\mathrm{pol/tor},k}(\theta, \phi)\sin(\theta)\,\mathrm{d}\theta \mathrm{d}\phi, k = r, \theta, \phi$.
We note that while $\langle B^2\rangle$ is a good proxy of the magnetic energy for the simulations, for the observations it is restricted to the net magnetic flux of the resolution elements.
The grey symbols represent the observed star sample from \cite{See2015}. Stars with masses $M_{\star} \geq 0.5\,\mathrm{M_{\odot}}$ are illustrated by grey squares and stars with masses $M_{\star} < 0.5\,\mathrm{M_{\odot}}$ by grey triangles. For a fair comparison between the simulations and observations, we present the simulations for two different resolutions of the large-scale field. The dark green circles contain all $\ell$-modes up to $\ell \leq 5$, which is comparable to ZDI-reconstructed maps of slowly rotating stars. The light green circles display the simulations for $\ell$-modes up to $\ell \leq 10$, which is comparable to ZDI-reconstructed maps of moderate rotators. Stellar magnetograms often include only $\ell \leq 5$ (e.g. \citealt{Morin2010, Vidotto2016a, Folsom2016}) or $\ell \leq 10$ modes (e.g. \citealt{Johnstone2014, Yadav2015}). In contrast high-resolution solar synoptic maps reach, e.g., $\ell \le 192$ \citep{DeRosa2012}. We add a dashed unity line in Fig.~\ref{fig:BtorBpol} for an easier identification of toroidal- and poloidal-dominated magnetic field topologies. 

It is clear from Fig.~\ref{fig:BtorBpol} (top) that the simulations fit entirely within the sample of the observed stars. They lie in the regime of the stars with masses above 0.5$\,\mathrm{M_{\odot}}$, which is reasonable as the simulations are based on the solar case. The simulated magnetic field topologies fall in the same area as the observed magnetic field topologies of e.g. the young solar-like stars HN\,Peg \citep{BoroSaikia2015} and $\varepsilon$\,Eri \citep{Jeffers2014}. The simulations aim to represent the solar-like stars not the observed stars with dominantly toroidal field or the M-dwarfs in the upper part of Fig.~\ref{fig:BtorBpol} (top). Both resolutions $\ell \leq 5$ and $\ell \leq 10$ are predominately poloidal, but the simulations restricted to $\ell \leq 5$ show on average a lower magnetic energy than the simulations with $\ell$-modes up to $\ell \leq 10$. This is expected as we are adding more energy by including more $\ell$-modes. 

%Fig. 2 bottom
In contrast to the observed stellar magnetic field topologies, the simulated magnetic field topologies provide information about both the large- and small-scale field. We study the magnetic field topology using the spherical harmonics $\ell = 1-28$, corresponding to length scales $\theta = 180^{\circ}$ to $\theta \approx 4.7^{\circ}$, where the solar simulation is fully-resolved. In Figure~\ref{fig:BtorBpol} (bottom) we over-plot the observed stars by the cumulative $\ell$-modes $\ell = 1$ to $\ell \leq 28$ of the simulations. We do not compare the simulations for different $\ell$-modes directly with the observations, as they have different resolutions. For a direct comparison between the simulations and the observations see Fig.~\ref{fig:BtorBpol} (top). The different $\ell$-modes in Fig.~\ref{fig:BtorBpol} (bottom) are colour-coded and divided in three regimes: the dipolar modes ($\ell = 1$) are illustrated by blue circles, the quadrupolar modes ($\ell \leq 2$) by orange circles and the higher $\ell$-modes ($\ell \leq 3, \ldots , \ell \leq 28$) by greenish circles, where the colour gets lighter with higher $\ell$-modes. By adding more and more $\ell$-modes, the magnetic energy increases in both the toroidal and poloidal fields and the spread in values decreases. The scatter of the simulations for a fixed resolution (equal coloured circles in Fig.~\ref{fig:BtorBpol}) is not larger than the spread of the observed stars. 

The dipolar modes $\ell = 1$ (blue circles) are highly poloidal and show the widest spread. However, the majority of the $\ell = 1$ simulations lie far from the unity line and appear as a classical dipole containing strong poloidal fields. Adding the $\ell \leq 2$ modes (orange circles) shifts all simulations by approximately one magnitude to higher toroidal magnetic energies. The quadrupolar modes $\ell=2$ are the most toroidal modes so that the magnetic field topologies for $\ell \leq 2$ may become dominantly toroidal as indicated by a few orange circles lying above the unity line. The inserts in Fig.~\ref{fig:BtorBpol} (bottom) show the poloidal (blue) and toroidal (red) field of an axisymmetric dipole and quadrupole mode. A strong toroidal $\ell = 2, m = 0$ quadrupole mode shows a polarity reversal across the equator similar to the emerged bipole pattern in Fig.~\ref{fig:SurfaceMaps} (bottom row).
As more $\ell$-modes are added, the magnetic field topologies become poloidal again (greenish circles). The higher cumulative $\ell$-modes ($\ell \leq 3-28$) show a fixed ratio between the magnetic energy stored in the toroidal and poloidal field that can be described by the power-law $\langle B^2_{\mathrm{tor}}\rangle \propto \langle B^2_{\mathrm{pol}}\rangle^{0.75 \pm 0.02}$, which was determined by a least squares best fit. The magnetic energy that is added by new modes $\ell \gtrsim 25$ asymptotically decreases to negligibly small numbers. The solar simulation captures $\langle B^2_{\ell \leq 5}\rangle / \langle B^2_{\mathrm{tot}}\rangle = 5.4\,\%$ or $\langle B^2_{\ell \leq 10}\rangle / \langle B^2_{\mathrm{tot}}\rangle = 21.6\,\%$ of the total energy, i.e. mean squared flux density.

% Fig 3: Axisym vs Toroidal
\begin{figure}
  	\includegraphics[width=\columnwidth]{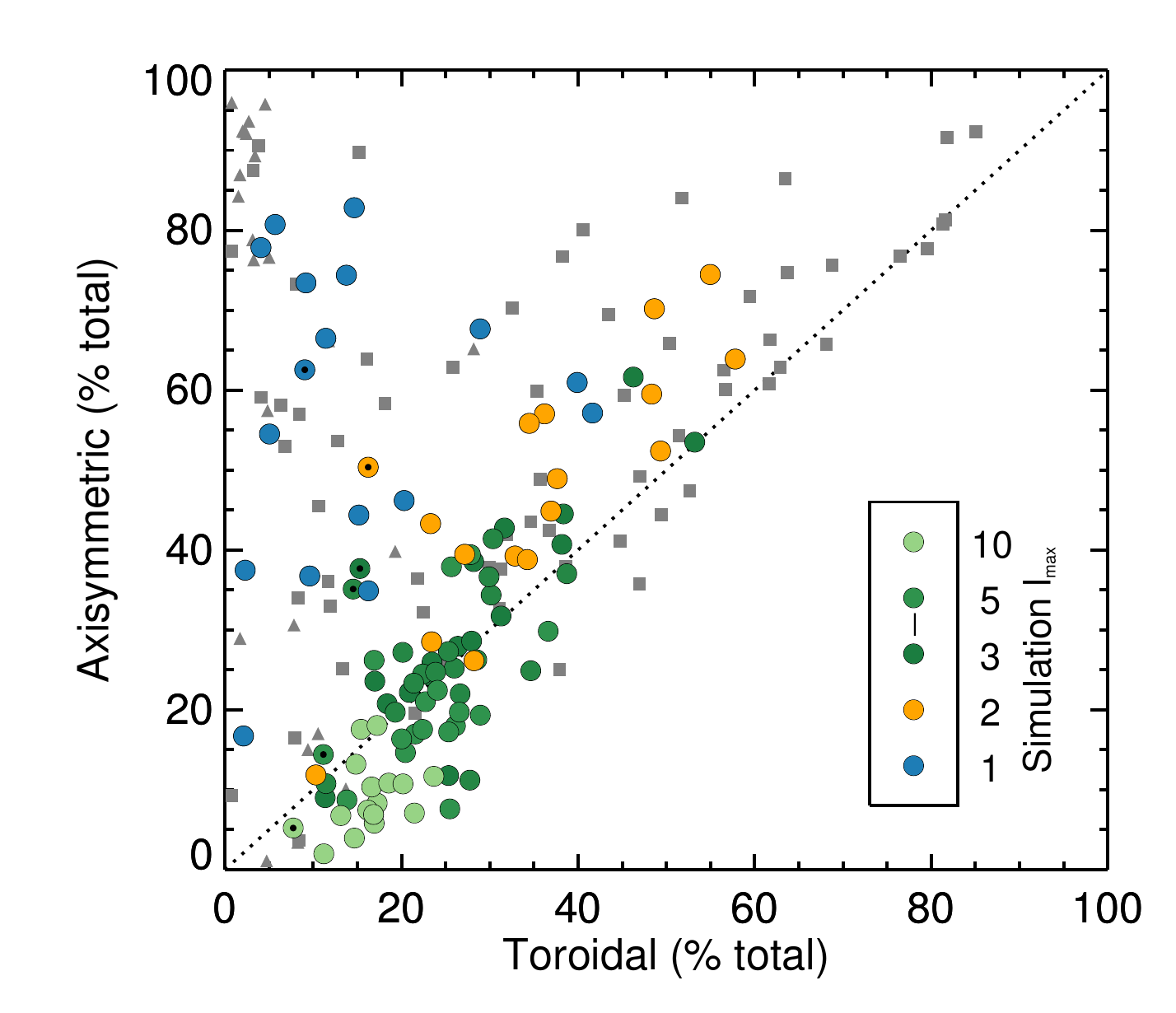}
    \caption{Comparison between simulations and observations of the percentage of axisymmetric and toroidal fields. The same format as in Fig.~\ref{fig:BtorBpol} is used.}
    \label{fig:AxiTor}
\end{figure}

\subsection{Axisymmetric vs toroidal fields}
\label{subsec:axitor}

Secondly, we analyse the axisymmetric component of the magnetic field topology. \cite{See2015} plotted the percentage of the total magnetic energy stored in the axisymmetric field against the percentage of the total magnetic energy stored in the toroidal field, \cite[Fig.~5]{See2015}. Figure~\ref{fig:AxiTor} here includes the results from the simulations of \cite{Gibb2016} in the same format as in Fig.~\ref{fig:BtorBpol}. The simulated sample is displayed by coloured circles for the different $\ell$-modes $\ell =1, \ell \leq 2, \ldots, \ell \leq 5$ and $\ell \leq 10$, where the colour scheme is the same as in Fig.~\ref{fig:BtorBpol}. 

The simulations cover the same parameter space as the observations in Fig.~\ref{fig:AxiTor}. For a fair comparison between the observed and the simulated stars, one has to focus on the green and light green circles representing the $\ell \leq 5$ and $\ell \leq 10$ modes. These modes show a small fraction of toroidal field and a comparably low fraction of axisymmetric field. In comparison, a value of <\,10\,\% toroidal field was determined for the Sun during CR2109 (Carrington rotation 2109 = 2011/04/12 - 2011/05/09) by \cite{Vidotto2016a}. Furthermore, we discover a trend with $\ell$-modes for the simulations. The dipolar component ($\ell =1$, blue circles) shows the biggest spread in the axisymmetric field and has in general low toroidal fields (similar to Fig.~\ref{fig:BtorBpol}~bottom). The quadrupolar component ($\ell \leq 2$, orange circles) displays the highest fraction of toroidal field of all modes (as seen in Fig.~\ref{fig:BtorBpol}~bottom). Additionally, strong toroidal fields are mainly axisymmetric fields. For the higher $\ell$-modes both the toroidal and axisymmetric fields decrease along the unity line until they saturate with 5-20\,\% toroidal fields and 0-5\,\% axisymmetric fields.

\section{Summary and Conclusions}

We have compared the magnetic field topologies of a sample of simulated stars modelled by \cite{Gibb2016} with the sample of observed stars analysed by \cite{See2015}. For both samples we focused on the magnetic energy stored in the poloidal and toroidal component and the fraction of axisymmetric fields. For a direct comparison between the simulations and observations we filtered out the small-scale field in the simulations using the spherical harmonic decomposition described by \cite{Vidotto2016a} to account for the difference in resolution between these samples. 
Additionally, we analysed the magnetic field topologies of the simulations for a larger range of surface scale sizes.
We discovered, that:
\begin{itemize}
\item The large-scale magnetic field topologies of the simulations fit into the parameter space covered by the solar-like stars within the observed sample. They do not, however, fit the stars with dominantly toroidal field or the M-dwarfs.
\item We identify for the simulations three different types of behaviour for $\ell =1, \ell \leq 2$, and all higher modes.
\item The dipolar component $\ell = 1$ of the simulations is mainly poloidal, whereas the quadrupolar component $\ell \leq 2$ displays the highest toroidal field fraction of all $\ell$-modes. Both components show a large spread in their axisymmetric fields but strong toroidal fields are strongly axisymmetric. 
\item The magnetic energies for the higher $\ell$-modes follow the power-law $\langle B^2_{\mathrm{tor}}\rangle \propto \langle B^2_{\mathrm{pol}}\rangle^{0.75 \pm 0.02}$. The highest $\ell$-modes add less and less energy to the total field until their magnetic energy become negligible. While increasing $\ell$-modes, the field becomes less axisymmetric and less toroidal.
\item The polarity pattern of the toroidal field of the emerging bipoles is noticeable through all $\ell$-modes down to $\ell \leq 2$. 
\end{itemize}
These results indicates that the global non-potential evolution model of \cite{Mackay2006} and applied by \cite{Gibb2016} captures key magnetic features of the solar-like stars in the observed sample through the process of differential rotation, meridional flow, surface diffusion and magnetic flux emergence.

Moreover, the magnetic field topologies of the simulations themselves display trends with increasing differential rotation and flux emergence rate. We will investigate this in a separate paper, where we study the influence of certain stellar properties, e.g., differential rotation, flux emergence rate, and meridional flow, on the magnetic field topologies at different length scales.

\section*{Acknowledgements}

The authors thank the referee Jeffrey Linsky for his constructive comments and suggestions. LTL acknowledges support from the Scottish Universities Physics Alliance (SUPA) prize studentship and the University of St Andrews Higgs studentship. MMJ and VS acknowledge a Science \& Technology Facilities Council (STFC) postdoctoral fellowship.

%%%%%%%%%%%%%%%%%%%%%%%%%%%%%%%%%%%%%%%%%%%%%%%%%%

%%%%%%%%%%%%%%%%%%%% REFERENCES %%%%%%%%%%%%%%%%%%

% The best way to enter references is to use BibTeX:

\bibliographystyle{mnras}
\bibliography{Lehmann2016MgFieldTopology3} % if your bibtex file is called example.bib

%%%%%%%%%%%%%%%%%%%%%%%%%%%%%%%%%%%%%%%%%%%%%%%%%%

%%%%%%%%%%%%%%%%% APPENDICES %%%%%%%%%%%%%%%%%%%%%

\appendix

\section{The poloidal and toroidal field}

Following \cite{Donati2006a} and \cite{Vidotto2016a}, we decompose the poloidal and toroidal field as:

\begin{align}
B_{\mathrm{pol}, r}(\theta, \phi) &\equiv B_{r}(\theta, \phi) = \sum_{\ell m} \alpha_{\ell m} P_{\ell m} e^{im\phi}, \\
B_{\mathrm{pol},\theta}(\theta,\phi) &= \sum_{\ell m} \beta_{\ell m} \frac{1}{\ell+1} \frac{\mathrm{d}P_{\ell m}}{\mathrm{d}\theta} e^{im\phi}, \\
B_{\mathrm{pol}, \phi}(\theta, \phi) &= - \sum_{\ell m} \beta_{\ell m} \frac{im P_{\ell m} e^{im\phi}}{(\ell + 1) \sin \theta},
\end{align}

\begin{align}
B_{\mathrm{tor}, r}(\theta, \phi) &= 0, \\
B_{\mathrm{tor},\theta}(\theta,\phi) &= \sum_{\ell m} \gamma_{\ell m} \frac{im P_{\ell m} e^{im\phi}}{(\ell + 1) \sin \theta}, \\
B_{\mathrm{tor}, \phi}(\theta, \phi) &=  \sum_{\ell m} \gamma_{\ell m} \frac{1}{\ell+1} \frac{\mathrm{d}P_{\ell m}}{\mathrm{d}\theta} e^{im\phi},
\end{align}
so that $\vec{B}_{\mathrm{pol}} + \vec{B}_{\mathrm{tor}} = \vec{B}$ \footnote{The radial field points outwards, the meridional ($\theta$) field increases from north to south with colatitude and the azimuthal field ($\phi$) increases in the direction of the rotation with longitude.}. The coefficients $\alpha_{\ell m}, \beta_{\ell m}, \gamma_{\ell m}$ characterise the magnetic field and $P_{\ell m} \equiv c_{\ell m}P_{\ell m}(\cos \theta)$ is the associated Legendre polynomial of mode $\ell$ and order $m$, where $c_{\ell m}$ is a normalization constant:
\begin{equation}
c_{\ell m} = \sqrt{\frac{2\ell+1}{4\pi}\frac{(\ell - m)!}{(\ell + m)!}}.
\end{equation}
The sums run from $1\leq \ell \leq \ell_{\mathrm{max}}$ and $\vert m \vert \leq \ell$, where $\ell_{\mathrm{max}}$ is the maximum mode of the spherical harmonic decomposition. The axisymmetric modes are selected by $m=0$. Otherwise, we sum over all $m$ for the magnetic field of a given mode $\ell$.

%%%%%%%%%%%%%%%%%%%%%%%%%%%%%%%%%%%%%%%%%%%%%%%%%%

% Don't change these lines
\bsp	% typesetting comment
\label{lastpage}
\end{document}